\documentclass[final,twocolumn,5p]{elsarticle}
\usepackage{multirow}
\usepackage{color}
\usepackage{graphics} 
\usepackage{rotating}
\usepackage{eqparbox}
\usepackage{graphics}
\usepackage{colortbl} 
 \usepackage{mathptmx} \usepackage[scaled=.90]{helvet} \usepackage{courier}
\usepackage{balance}
\usepackage{picture}
\usepackage{algorithm}
\usepackage{algorithmicx}
\usepackage{algpseudocode}
\usepackage[export]{adjustbox}
\renewcommand{\footnotesize}{\scriptsize}
\definecolor{lightgray}{gray}{0.8}
\definecolor{darkgray}{gray}{0.6}

\definecolor{Gray}{gray}{0.95}
\definecolor{LightGray}{gray}{0.975}
\newcommand{\bi}{\begin{itemize}[leftmargin=0.4cm]}
\newcommand{\ei}{\end{itemize}}
\newcommand{\be}{\begin{enumerate}}
\newcommand{\ee}{\end{enumerate}}
\newcommand{\tion}[1]{\S\ref{sect:#1}}
\newcommand{\fig}[1]{Figure~\ref{fig:#1}}
\newcommand{\tab}[1]{Table~\ref{tab:#1}}
\newcommand{\eq}[1]{Equation~\ref{eq:#1}}

\usepackage[shortlabels]{enumitem}  
\usepackage{url}
\begin{document}
\begin{frontmatter}
\title{ Tuning for Software Analytics: is it Really Necessary?}
\author{Wei Fu\corref{cor1}}
\ead{wfu@ncsu.edu}
\author{Tim Menzies\corref{cor1}}
\ead{tim.menzies@gmail.com}
\author{Xipeng Shen}
\ead{ xshen5@ncsu.edu}
\cortext[cor1]{Corresponding author: Tel:19195345251(Wei)}
\address{Department of Computer Science, North Carolina State University, Raleigh, NC, USA}

% % \numberofauthors{1}
% \author{Wei Fu \and Tim Menzies \and Xipeng Shen}
% \institute{North Carolina State University, Raleigh, NC, USA
%       Wei Fu \email{w}}
% % \email{fuwei.ee \and tim.menzies@gmail.com \and xshen5@ncsu.edu }

% \thispagestyle{plain}
% \pagestyle{plain}
\begin{abstract}
\textbf{Context:} Data miners have been widely used in
software engineering to, say,  generate defect predictors from static code measures. Such static code defect predictors perform well compared to manual methods, and they are easy to use and useful to use.  But 
one of the ``black arts'' of data mining is setting the tunings that control the miner. \\
\textbf{Objective:} We seek  simple, automatic, and very effective  method for finding those tunings.\\
\textbf{Method:} For each experiment with different data sets (from open source JAVA systems), we ran differential evolution as an optimizer to explore the tuning space
(as a first step) then tested the tunings
using hold-out data.\\
\textbf{Results:} Contrary to our prior expectations, we found these tunings
 were remarkably simple: it only required tens, not thousands, of attempts
 to obtain very good results. For example, when  learning software defect predictors, this  method can quickly
 find tunings  that  alter detection  precision
 from 0\% to 60\%. \\
\textbf{Conclusion:} Since (1)~the improvements are so large, and (2)~the tuning is so simple, we need 
to change  standard methods in software analytics.
At least for defect prediction, 
it is no longer enough to just run a data miner and present the result
{\em without}  conducting a tuning optimization study.
The implication for other kinds of  analytics is now  an open and pressing issue.

\end{abstract}
\end{frontmatter}

% % A category with the (minimum) three required fields
% \vspace{1mm}
% \noindent
% {\bf Categories/Subject Descriptors:} 
% D.2.8 [Software Engineering]: Product metrics;
% I.2.6 [Artificial Intelligence]: Induction

\vspace{1mm}
\noindent
{\bf Keywords:} defect prediction, CART, random forest,
differential evolution,
search-based software engineering.
%  \maketitle 
\pagenumbering{arabic} %XXX delete before submission

\section{Introduction}
 
In the $21^{st}$ century, it is  impossible
to manually browse all  available software project
data. The PROMISE repository of SE data has grown to 200+ projects~\cite{promise15}
and this is just one of over a dozen open-source repositories
that are readily available to researchers~\cite{rod12}.
For example, at the time of this writing (Jan 2016), our web searches show that Mozilla Firefox has over 1.1 million bug reports, and platforms such as GitHub host over 14 million projects.

Faced with this data overload,
researchers in empirical SE
use  data miners  to generate 
{\em defect predictors from static code measures}.
Such   measures can be
automatically extracted from the code base, with very little effort
even for very large software systems~\cite{nagappan05}. 

One of the ``black arts'' of data mining is setting the tuning
parameters that control  the choices within a data miner.
Prior to this work, our intuition was that tuning would change the behavior or a data miner, to some degree. Nevertheless, we rarely tuned our  defect predictors 
since we reasoned
that a data miner's default tunings have been well-explored by the developers of those algorithms (in which case
tuning would not lead to large performance improvements).
Also, we suspected that
tuning would take so long time and be so CPU intensive that the benefits gained   would not be worth effort.

The results of this paper show that the above points are
false since, at least for
defect prediction from  code attributes:
\be
\item
Tuning  defect predictors is {\em remarkably simple};
\item
And can {\em dramatically improve the performance}. 
\ee
Those results were found by   exploring six research questions:
\bi
\item RQ1: {\em Does   tuning    improve the performance scores of a predictor?} We will show below
 examples of truly dramatic improvement:
 usually by 5 to 20\% and often by much more (in one extreme case,  precision improved from 0\% to 60\%).
\item RQ2: {\em Does tuning change conclusions on what learners are better than others?} 
Recent SE papers~\cite{lessmann2008benchmarking,hall11} claim that some learners are better than others. 
Some of those conclusions are completely changed by tuning. 
\item RQ3: {\em Does tuning change conclusions about what factors are most important in software engineering?} Numerous recent SE papers (e.g.~\cite{bell2013limited,rahman2013how,me02k,Moser:2008,zimmermann2007predicting,%
herzig2013predicting}) use data miners to conclude that {\em this}
is more important than {\em that} for reducing software project defects.
Given the  tuning results of this paper, we show that such conclusions need to be revisited.
\item  RQ4: {\em Is tuning easy?} We show that one of the simpler multi-objective optimizers
(differential evolution~\cite{storn1997differential}) works very well for tuning defect predictors. 
\item RQ5: {\em Is tuning impractically slow?} We achieved dramatic improvements in the performance scores
of our data miners in less than 100 evaluations (!); i.e., very
quickly.
\item RQ6: {\em Should data miners be used ``off-the-shelf'' with their default tunings?} 
For defect prediction from static code measures, our answer is an emphatic ``no'' (and
the implication for other kinds of  analytics is now an open and urgent question).
\ei
Based on our answers to these  questions,  we strongly advise that:
\bi
\item
Data miners should not be used ``off-the-shelf'' with default tunings.
\item
Any future paper on defect prediction should include a 
tuning study. Here, we have found  an algorithm called differential
evolution to be a useful method for conducting such
tunings.
\item
Tuning needs to be repeated
whenever data or goals are changed.
Fortunately, the cost of finding good tunings is not excessive since, at least for
static code defect predictors, tuning is easy and fast.
\ei

%Before beginning, we digress for one  clarification.
%This paper is {\em not} arguing that
%software analytics is somehow wrong-headed or misguided.
%In the age of the Internet and global access to software engineering data,
%there exists the  problem of information overload. {\em Something} must be %done to
%allow analysts to make conclusions via an automatic analysis over a lot of %data.
%The results of this paper is that for a particular local context
%(a specific data set and a specific goal) there exists  
%methods for optimizing the conclusions reached in that context.  
%Those conclusions
%may not generalize to other contexts but this  is not a council for despair. 
%As shown here, 
%there  exists general methods for finding
%local conclusions in a particular context. Further,
%those
%methods are  very simple to implement and very fast to execute.

\section{Preliminaries}

\subsection{Tuning: Important and Ignored}
This section argues that   tuning is an under-explored software analytics-- particularly in the apparently  well-explored field of defect prediction.

In  other fields, the impact of tuning is well understood~\cite{Bergstra2012}. 
Yet issues of tuning  are rarely or poorly addressed
in the defect prediction literature.
When we tune a data miner, what we are really doing is changing how a learner applies
its heuristics. This means tuned data miners use different heuristics, which means they ignore different possible models, which means they return different models; i.e.  {\em how} we learn changes {\em what} we learn.

Are the impacts of tuning addressed in the defect prediction literature?
To answer that question,  in Jan 2016 we searched scholar.google.com for the conjunction of ``data mining" and ``software engineering" and  ``defect prediction"
(more details can be found at https://goo.gl/Inl9nF).
After sorting by the citation count and discarding the non-SE papers (and those without a pdf link), we read over this sample
of  50 highly-cited SE defect prediction papers. 
What we found in that sample was that few authors
acknowledged the impact of tunings (exceptions:~\cite{Gao:2011,lessmann2008benchmarking}).
Overall,  80\% of papers in our sample {\em did not} adjust
the ``off-the-shelf'' configuration of the data miner (e.g.~\cite{me07b,Moser:2008,Elish2008649}). Of the remaining papers:
\bi
\item
Some papers in our sample  explored   data super-sampling~\cite{4271036} or data sub-sampling techniques via  automatic methods (e.g. ~\cite{Gao:2011,me07b,4271036,Kim:2011}) 
or via some domain principles (e.g. ~\cite{Moser:2008,Nagappan:2008,Hassan:2009}).
As an example of the latter, Nagappan et al.~\cite{Nagappan:2008} checked if metrics related to organizational structure were relatively more powerful for predicting software defects. 
However, it should be noted that  these studies varied the input data but
not the   ``off-the-shelf''   settings of the data miner.
\item
A few other papers did acknowledge that one data miner may not be appropriate
for all data sets.  Those papers tested  different  
``off-the-shelf'' data miners on the same data set.
For example, Elish et al.\cite{Elish2008649}  compared support vector
machines to other data miners for the purposes of defect prediction. SVM's execute via a ``kernel function'' which should be specially selected for different data sets and
the Elish et al. paper  makes no mention of any SVM tuning study.  
To be fair to Elish et al., we hasten to add that we
ourselves have  published
papers using ``off-the-shelf'' tunings~\cite{me07b} since,
prior to this paper it was unclear to us how to effectively
navigate the large space of possible tunings.
\ei
Over our entire sample, there was only  one paper that conducted a somewhat extensive tuning study.
Lessmann et al.\cite{lessmann2008benchmarking} tuned parameters for some of their algorithms using  a {\em grid search}; i.e. divide all $C$ configuration
options into $N$ values, then try all   $N^C$ combinations.
This is a slow approach-- we have explored grid search for 
defect prediction and found it takes days to terminate~\cite{me07b}.
Not only that, we found that grid search can miss
important optimizations~\cite{baker07}.
Every grid has ``gaps'' between each grid division which means
that a supposedly rigorous grid search can still miss
important configurations~\cite{Bergstra2012}. 
Bergstra and Bengio~\cite{Bergstra2012} comment that for most data sets only a few of the tuning parameters really matter-- which means that
much of the runtime associated with grid search is actually wasted.
Worse still, Bergstra and Bengio  comment that 
the 
important tunings are   different   for different
data sets-- a 
 phenomenon that makes grid search a poor choice for configuring data mining
 algorithms for new data sets.

Since the Lessmann et al. paper, much progress has been made in 
configuration algorithms
and we can now report that  {\em finding useful tunings is very easy}.
This result is both novel and unexpected.
A standard run of grid search (and other  evolutionary algorithms)
is  that optimization requires   thousands,
if not millions, of evaluations.  However, in a result that we found startling, that  {\em differential evolution} (described below) can find useful settings for learners generating defect predictors
in less than 100 evaluations (i.e. very quickly).
Hence,   the ``problem'' (that
tuning changes the conclusions) is really
an exciting opportunity. At least for defect prediction,
 learners are very   amenable to tuning. Hence, 
 they are  also very
amenable to significant performance improvements. Given the low
number of evaluations required, then we assert that tuning
  should be standard practice
for anyone building defect predictors.

 \subsection{You Can't Always Get What You Want}\label{sect:goals}
 
 Having made the case that tuning needs to be explored more,
 but before we get into the technical details of this
 paper, this section discusses some
 general matters about setting goals during tuning
 experiments.
 
 This paper characterizes tuning as an optimization problem (how to change the settings on the learner
 in order to best improve the output).
With such optimizations,  it is not always possible to optimize for all goals at the same time.
For example, the following text does not
show results for tuning on recall
or false alarms since optimizing {\em only} for those goals can lead
to some undesirable side effects:
\bi
\item
{\em Recall} reports the percentage of  predictions that are actual examples of  what we are looking for.
When we tune for {\em recall}, we can achieve near
100\% recall-- but at the cost of a near 100\% false alarms.
\item
{\em False alarms} is the percentage of other examples that are reported  (by the learner)
to be part of the targeted examples.
When we tune for {\em false alarms}, we 
can achieve near zero percent false alarm rates by effectively turning off
the detector (so the  recall falls to nearly zero).
\ei
Accordingly,  this paper  explores performance measures that comment on all 
target classes: see the  precision and ``F'' measures discussed below: see {\em Optimization Goals}.
That said, we are sometimes asked what good is a learner if it optimizes for (say) precision
at the expense of (say) recall. 

Our reply is that software engineering is a very diverse enterprise
and that different kinds of development need to optimize for different goals
(which may not necessarily be ``optimize for recall''):
\bi
\item
Anda, Sjoberg and Mockus are concerned with {\em reproducibility}  and so
assess their models using the the ``coefficient of variation'' ($CV = \frac{stddev}{mean}$) ~\cite{anda09}.
\item
Arisholm~\&~Briand~\cite{arisholm06},  Ostrand \& Weyeuker~\cite{ostrand04} and Rahman et al.~\cite{rahman12} are concerned with reducing the work load associated with someone
else reading a learned model, then applying it. Hence, they assess their models using  {\em reward}; i.e.   the fewest lines of code
  containing the most bugs.
\item
Yin et al. are concerned about
 {\em incorrect bug fixes}; i.e. those that require subsequent work in order to complete the bug fix.
These bugs occur  when (say) developers try to fix parts of the code
where they have very little experience~\cite{yin11}.  Hence, they assess a learned
model using a measure that selects for  the most number of bugs in regions that {\em the most programmers have worked with before}.
\item
For safety critical applications, high false alarm rates are  acceptable if the cost
of overlooking  critical issues outweighs the inconvenience of   inspecting a few more
modules. 
\item
When rushing a product to market,  there is a business case to 
avoid the extra rework associated with false alarms.  In that business context, 
managers might be willing to lower the recall somewhat in order to minimize the false alarms.
\item
When the second author worked with contractors at  NASA's software independent verification
and validation facility, he found  new contractors  
only reported issues that were most certainly important defects; i.e. they minimized
  false alarms even if that damaged their precision (since, they felt, 
it was better to be silent than wrong). Later on, once
those contractors had acquired a reputation of being insightful members of the team,
they improved their precision scores (even if it means some more false alarms).
\ei
Accordingly, this paper does not assume that (e.g.) minimizing false alarms is 
more important than maximizing   precision or recall. Such a determination 
depends on   business conditions.

Rather, what we can  show  examples where  changing  optimization goals can also change 
the conclusions made from that learner on that data. More generally, we caution that it is 
important not to overstate  empirical results from  analytics.
Those results need to be expressed {\em along with} the context within which they are
relevant (and by ``context'', we mean the optimization goal).

\subsection{Notes on Defect Prediction}

This section discusses defect prediction,
which is the particular
task explored by our optimizers.

Human programmers are clever, but flawed. Coding  adds functionality, but also defects.
Hence, software sometimes crashes (perhaps at the most awkward or dangerous moment) or delivers
the wrong functionality. For a very long list of software-related errors,
see  Peter Neumann's ``Risk Digest'' at catless.ncl.ac.uk/Risks.

Since programming inherently
introduces defects into  programs, it is important to test them before they're used.
Testing is expensive.
Software assessment budgets are finite
while assessment effectiveness increases 
exponentially with assessment effort.
For example, for  black-box testing methods,
a {\em linear} increase
in the confidence $C$ of finding  defects
can take {\em exponentially} more effort:
\bi
\item
A randomly selected 
input to a program will find a fault with probability $p$.
\item
After $N$ random black-box tests, the chances of the inputs 
not revealing any fault 
is $(1-p)^N$. 
\item
Hence, the chances $C$ of seeing the fault is $1-(1-p)^N$
which can be rearranged to 
 $N(C,p)=log(1 -
C)/log(1-p)$. 
\item
For example, $N(0.90,10^{-3})=2301$
but $N(0.98,10^{-3})=3901$; i.e. nearly double the number of tests.
\ei
Exponential costs quickly exhaust finite resources so
standard practice is to apply the best
available  methods on code sections that seem   most critical. 
But 
any method that focuses on parts of the code
can blind us to defects in other areas. Some  {\em
lightweight sampling policy} should be used to explore the rest of the system.  This
sampling policy will always be incomplete.
Nevertheless, it is the only option when
resources prevent a complete assessment of everything.

One such lightweight sampling policy is defect predictors learned from static code attributes.
Given software described in the attributes of \tab{ck},   data miners can
learn where the probability of software defects is highest.

The rest of this section argues that such defect predictors are   {\em easy to
use}, {\em widely-used}, and {\em useful} to use.

\begin{table*}[t]
\renewcommand{\baselinestretch}{0.8}\begin{center}
{\scriptsize
\begin{tabular}{c|l|p{4.7in}}
amc & average method complexity & e.g. number of JAVA byte codes\\\hline
avg\_cc & average McCabe & average McCabe's cyclomatic complexity seen
in class\\\hline
ca & afferent couplings & how many other classes use the specific
class. \\\hline
cam & cohesion amongst classes & summation of number of different
types of method parameters in every method divided by a multiplication
of number of different method parameter types in whole class and
number of methods. \\\hline
cbm &coupling between methods &  total number of new/redefined methods
to which all the inherited methods are coupled\\\hline
cbo & coupling between objects & increased when the methods of one
class access services of another.\\\hline
ce & efferent couplings & how many other classes is used by the
specific class. \\\hline
dam & data access & ratio of the number of private (protected)
attributes to the total number of attributes\\\hline
dit & depth of inheritance tree &\\\hline
ic & inheritance coupling &  number of parent classes to which a given
class is coupled (includes counts of methods and variables inherited)
\\\hline
lcom & lack of cohesion in methods &number of pairs of methods that do
not share a reference to an instance variable.\\\hline
locm3 & another lack of cohesion measure & if $m,a$ are  the number of
$methods,attributes$
in a class number and $\mu(a)$  is the number of methods accessing an
attribute, 
then
$lcom3=((\frac{1}{a} \sum_j^a \mu(a_j)) - m)/ (1-m)$.
\\\hline
loc & lines of code &\\\hline
max\_cc & maximum McCabe & maximum McCabe's cyclomatic complexity seen
in class\\\hline
mfa & functional abstraction & number of methods inherited by a class
plus number of methods accessible by member methods of the
class\\\hline
moa &  aggregation &  count of the number of data declarations (class
fields) whose types are user defined classes\\\hline
noc &  number of children &\\\hline
npm & number of public methods & \\\hline
rfc & response for a class &number of  methods invoked in response to
a message to the object.\\\hline
wmc & weighted methods per class &\\\hline
\rowcolor{lightgray}
defect & defect & Boolean: where defects found in post-release bug-tracking systems.
\end{tabular}
}
\end{center}
\caption{OO measures used in our defect data sets.}\label{tab:ck}
\end{table*}

{\em Easy to use:} Static code attributes can be automatically collected, even for very large systems~\cite{nagappan05}.
Other methods, like  manual code reviews, are far slower and far more labor-intensive.
For example, depending on the review methods, 8 to 20 LOC/minute can be
inspected and this effort repeats for all members of the review team,
which can be as large as four or six people~\cite{me02f}. 
{\em Widely used:}  Researchers and industrial practitioners  use static attributes to guide software 
quality predictions.
 Defect prediction models have been reported
  at Google~\cite{lewis13}.
Verification and validation (V\&V) textbooks
\cite{rakitin01} advise using static code complexity attributes
to decide which modules are worth manual inspections.

{\em Useful:}
Defect predictors often  find the location of  70\% (or more)
of the defects in code~\cite{me07b}.
Defect predictors have some level of generality:
predictors learned at NASA~\cite{me07b} have also been found useful elsewhere
(e.g. in Turkey~\cite{tosun10,tosun09}).
The success of this method in  predictors in finding bugs is   markedly
higher than other currently-used
industrial
methods such as manual code reviews. For example, 
a  panel at {\em IEEE Metrics
2002}~\cite{shu02} concluded that manual software  reviews can find ${\approx}60\%$ 
of defects.
In another work, 
Raffo documents the typical    defect detection capability of
industrial review methods:   around 50\%
 for full Fagan inspections~\cite{fagan76} to
21\% for less-structured inspections.

Not only do static code defect predictors perform well compared to manual methods,
they also are competitive with certain automatic methods.
A recent study at ICSE'14, Rahman et al.~\cite{rahman14:icse} compared
(a) static code analysis tools FindBugs, Jlint, and Pmd and (b)
static code defect predictors
(which they called ``statistical defect prediction'') built using logistic regression.
They found  no significant differences in the cost-effectiveness
of these  approaches. Given this equivalence, it is significant to note that 
static code defect prediction can be quickly adapted to new languages by building lightweight
parsers that find   information like \tab{ck}. The same is not true for   static code analyzers-- these need  extensive modification before they can be used on new
languages.

 \subsection{Notes on Data Miners}
 
There are several ways to make defect predictors
using  CART~\cite{brieman00}, Random Forest~\cite{breiman84}, 
 WHERE~\cite{menzies2013local} and LR (logistic regression).
For this study, we use CART, Random Forest and LR versions  from 
SciKitLearn~\cite{scikit-learn} and
WHERE, which is available from
github.com/ai-se/where. 
 We use  these algorithms for the following reasons.
 
CART and Random Forest were mentioned in
a recent IEEE TSE paper by Lessmann et al.~\cite{lessmann2008benchmarking} that compared 22  
learners for  defect prediction.
That study ranked  CART  worst  and Random Forest as best.
In a demonstration of the impact of tuning,
this paper shows  we can {\em refute} the conclusions of  Lessmann et al.
in the sense that, after tuning,
CART
performs just as well as
 Random Forest.

LR was  mentioned by Hall et al.~\cite{hall11}
as usually being as good or better as more complex learners (e.g.
Random Forest). In a finding that endorses the Hall et al. result,
we show that untuned LR performs better than 
untuned Random Forest (at least, for the data sets studied here). However,
we will show that tuning raises doubts about the optimality of the
Hall et al. recommendation.

Finally,  this
 paper uses WHERE since, as shown below,
it offers an interesting case study on the benefits of tuning.

%%%%%%%%%%%%%%%% list of parameters%%%%%%%%%%%%%%%%%%%%%
\renewcommand\arraystretch{1.2}
\begin{table*}[t!]
\scriptsize
  \centering
	\begin{tabular}{|c|c|c|c|l|}
	\cline{1-5}
	\begin{tabular}[c]{@{}c@{}}Learner Name\end{tabular} & Parameters & Default &\begin{tabular}[c]{@{}c@{}}Tuning\\ Range\end{tabular}& 
\multicolumn{1}{c|}{Description} \\ \hline
 	\multirow{8}{*}{\begin{tabular}[c]{@{}c@{}}Where-based\\ Learner\end{tabular}}
%  	WHERE-based  Learner} 
	& threshold & 0.5 &[0.01,1]& The value to determine defective or not .\\ \cline{2-5} 
	& infoPrune & 0.33 &[0.01,1]& The percentage of features to consider for the best 
split to build its final decision tree. \\ \cline{2-5} 
	 & min\_sample\_split & 4& [1,10]& The minimum number of samples required to split an internal node of
its final  decision tree. \\ \cline{2-5} 
	 & min\_Size & 0.5 &[0.01,1]& \begin{tabular}[c]{@{}l@{}}Finds min\_samples\_leaf 
in the initial clustering tree using ${n\_samples}^ {min\_Size}$.
\end{tabular} \\ \cline{2-5} 
    & wriggle & 0.2 &[0.01, 1] & The threshold to determine which branch in  the initial clustering tree to be pruned\\ \cline{2-5}
	 & depthMin & 2 & [1,6]&The minimum depth of the initial clustering tree below which no pruning for the
clustering tree. \\ \cline{2-5} 
	 & depthMax & 10 &[1,20]& The maximum depth of the initial clustering tree. \\ \cline{2-5} 
	 & wherePrune & False &T/F& Whether or not to prune the initial clustering tree. \\ \cline{2-5}
	 & treePrune & True &T/F& Whether or not to prune the final decision tree. \\ \cline{2-5} 
\hline
\multirow{5}{*}{CART} & threshold & 0.5 &[0,1]& The value to determine defective or not. \\ \cline{2-5} 
	 & max\_feature & None &[0.01,1]& The number of features to consider when looking for the best 
split. \\ \cline{2-5} 

	 & min\_sample\_split & 2 &[2,20]& The minimum number of samples required to split an 
internal node. \\ \cline{2-5} 
	 & min\_samples\_leaf & 1 & [1,20]&The minimum number of samples required to be at a leaf 
node. \\ \cline{2-5} 
     & max\_depth & None & [1, 50]& The maximum depth of the tree. \\
     \cline{1-5}  
       \multirow{5}{*}{\begin{tabular}[c]{@{}c@{}}Random \\ Forests\end{tabular}}  & threshold & 0.5 & [0.01,1] & The value to determine defective or not. \\ 
\cline{2-5} 
	 & max\_feature & None &[0.01,1]& The number of features to consider when looking for the best 
split. \\ \cline{2-5} 
	 & max\_leaf\_nodes & None &[1,50]& Grow trees with max\_leaf\_nodes in best-first fashion. \\ \cline{2-5} 
	 & min\_sample\_split & 2 &[2,20]& The minimum number of samples required to split an 
internal node. \\ \cline{2-5} 
	 & min\_samples\_leaf & 1 &[1,20]&The minimum number of samples required to be at a leaf 
node. \\ \cline{2-5} 
	 &  n\_estimators & 100 & [50,150]&The number of trees in the forest.\\ \cline{2-5}
	 \hline 
Logistic Regression&\multicolumn{4}{c|}{This study uses untuned LR in order to check
a conclusion of~\cite{hall11}. }\\\hline

	\end{tabular}
    \caption {List of parameters tuned by this paper.}
\label{tab:parameters}
\end{table*}

\subsection{ Learners and Their Tunings}

Our learners use the tuning parameters of \tab{parameters}. This section describes those parameters.
The default parameters for CART and Random Forest are set by 
the SciKitLearn authors and the
default parameters for WHERE-based learner are set via our own expert judgement.
When we say a learner is used ``off-the-shelf'', we mean
that they use the defaults shown in \tab{parameters}. 

As to the value of those defaults, it could be argued that these defaults are 
not the best  parameters for practical defect prediction.
That said,  prior to this paper, two things were true:
\bi
\item 
Many data scientists in SE use the standard defaults
in their data miners, 
without   tuning (e.g.~\cite{me07b,Moser:2008,herzig2013predicting,zimmermann2007predicting}).
\item
The effort involved to adjust those tunings seemed so onerous, that
many researchers in this field were content to take our prior advice
of ``do not tune... it is just too hard''~\cite{me15:book1}.
\ei
As to why we used the "Tuning Range" shown in \tab{parameters}, and not some other ranges,
we note that (1)~those ranges included the defaults; (2)~the results shown below
show that by exploring those ranges,   we achieved large gains in the performance of our defect predictors.
This is not to say that {\em larger} tuning ranges might not result in {\em greater} improvements.
However, for the goals of this paper (to show that some tunings do matter), exploring
just these ranges shown in \tab{parameters} will suffice.

As to the details of these learners, LR is a parametric
modeling approach. Given $f = \beta_0 + \sum_i\beta_ix_i$,
where $x_i$ is some measurement in a data set, and $\beta_i$
is learned via regression, LR
converts that into a function $0 \le g \le 1$
using $g=1/\left(1+e^{- f}\right)$. This function reports how much
we believe in a particular class. 

CART, Random Forest, and WHERE-based learners are all  tree learners that divide a data set, then recur
on each split.
All these learners
generate numeric predictions which are converted
into binary ``yes/no'' decisions via \eq{yesno}.

\begin{equation}\label{eq:yesno}\scriptsize
\mathit{inspect}= \left\{
\begin{array}{ll}
d_i \ge T \rightarrow \mathit{Yes}\\
d_i <   T \rightarrow \mathit{No} ,
\end{array}\right.
\end{equation}
where $d_i$ is the number of observed issues and $T$
is some threshold defined by an engineering judgement; we use $T=1$.

The splitting process is controlled by numerous tuning parameters.
If data contains more than {\em min\_sample\_split}, then a split is attempted.
On the other hand, if a split contains no more than {\em min\_samples\_leaf}, then the recursion stops. CART and Random Forest use a 
user-supplied constant for this parameter while
WHERE-based learner firstly computes this parameter $m$={\em min\_samples\_leaf} from the size of the data
sets via  $m=\mathit{size}^\mathit{min\_size}$ to build an initial  clustering tree.
Note that WHERE builds {\em two} trees: the initial clustering tree (to find similar sets of data)
then a final decision tree (to learn rules that predict for each similar cluster).
A
frequently asked question is why does WHERE build two trees--
would not   a single tree suffice? The answer is, as shown below,  tuned WHERE's twin-tree approach 
generates very precise predictors.
As to the rest of WHERE's parameters, the 
 parameter  {\em min\_sample\_ split } controls the construction of the
final decision tree (so, for WHERE-based learner,
{\em min\_size} and {\em min\_sample\_split} are the parameters to be tuned).

These learners use different techniques to explore the splits:
\bi
\item
CART finds the attributes whose ranges contain rows with least variance in the number
of defects. If an attribute ranges $r_i$ is found in 
$n_i$ rows each with a  defect count variance of $v_i$, then CART seeks the attributes
whose ranges minimizes $\sum_i \left(\sqrt{v_i}\times n_i/(\sum_i n_i)\right)$.
\item
Random Forest    divides data like CART then  builds $F>1$  trees,
each time using some random subset of
the attributes. 
\item
When building the initial cluster tree, WHERE projects the data on to a dimension it synthesizes from the raw data using
a process analogous to principle component analysis~\cite{jolliffe2002principal}

% \footnote{
% PCA  synthesises  new
% attributes $e_i, e_2,...$
% that extends across the dimension of greatest  variance in the data  with attributes $d$.  
% This process  combines
% redundant  variables into a smaller set of variables  (so $e \ll d$) since those
% redundancies become (approximately) parallel lines
% in $e$ space. For all such redundancies \mbox{$i,j \in d$}, we 
% can ignore $j$ 
% since effects that change over $j$ also
% change in the same way over $i$.
% PCA is also useful for skipping over noisy variables from $d$-- these
% variables are effectively ignored since    they  do not contribute to the variance in the data.}.
WHERE  divides  at the median point of that projection.
On recursion,
this generates the initial clustering tree, the leaves of which are clusters of  very similar examples. After that, when building 
the final decision tree, WHERE pretends its clusters are ``classes'', then 
asks the InfoGain algorithm of the
Fayyad-Irani discretizer~\cite{FayIra93Multi}, to rank the attributes, where {\em infoPrune} is used.
WHERE's final decision tree generator then ignores everything except the top   {\em infoPrune} percent of the sorted
attributes.
\ei
Some tuning parameters are learner specific:
\bi
\item
{\em Max\_feature} is used by
CART and Random Forest to select the number of attributes
used to build one tree.
CART's default is to use all the attributes while 
Random Forest usually selects the square root of the number
of attributes.
\item
  {\em Max\_leaf\_nodes} is the upper bound on leaf notes generated in a 
  Random Forest.
\item {\em Max\_depth} is the upper bound on the depth of the CART tree.  
 \item
WHERE's  
tree generation will always split up to {\em depthMin} number of branches.
After that, WHERE will only split data if the mean performance scores of the two halves
is ``trivially small'' (where ``trivially small'' is set by the   {\em wriggle} parameter). 
\item
WHERE's   {\em tree\_prune} setting controls how   
WHERE prunes back superfluous parts of the final decision tree. 
If a decision sub-tree and its parent have the same 
majority cluster
(one that occurs most frequently), then if {\em tree\_prune} is enabled, we prune that decision sub-tree.
\ei

\subsection{Tuning Algorithms}

How should researchers select which optimizers to apply to tuning data miners?
Cohen~\cite{cohen95} advises comparing new 
 methods against the simplest possible alternative. 
Similarly, Holte~\cite{holte93} recommends using very simple  learners
 as a
kind of ``scout'' for a  preliminary analysis of a data
set (to check if that data really requires a more
complex analysis).
Accordingly,
to find our ``scout'',  we used engineering judgement to sort  candidate algorithms from simplest to  complex. For
example, here is a list of optimizers used widely in research:
{\em 
simulated annealing}~\cite{fea02a,me07f};
 various {\em genetic algorithms}~\cite{goldberg79} augmented by
techniques such as {\em differential evolution}~\cite{storn1997differential}, 
{\em tabu search} and {\em scatter search}~\cite{Glover1986563,Beausoleil2006426,Molina05sspmo:a,4455350};
{\em particle swarm optimization}~\cite{pan08}; 
numerous {\em decomposition} approaches that use
    heuristics to decompose the total space into   small problems,   then apply a
    {\em response surface methods}~\cite{krall15,Zuluaga:13}.
Of these,  the simplest are simulated annealing (SA)  and 
differential evolution (DE), each of which can be coded in less than a page of some high-level scripting language. Our reading of the current literature is that there are more  advocates for
differential evolution than
  SA. For example,  Vesterstrom and Thomsen~\cite{Vesterstrom04} found DE to be competitive with 
   particle swarm optimization and other GAs. 
   
DEs have been applied before for   parameter tuning (e.g. see~\cite{omran2005differential, chiha2012tuning}) but this is the first time they have been applied to
optimize defect prediction from static code attributes.  
The pseudocode for differential evolution is shown in Algorithm~\ref{alg:DE}.
In the following description, 
    superscript numbers denote lines in that pseudocode.

\begin{algorithm}[!t]

\scriptsize
\begin{algorithmic}[1]
\Require $\mathit{np} = 10$, $f=0.75$, $cr=0.3$, $\mathit{life} = 5$, $\mathit{Goal} \in \{\mathit{pd},f,...\}$
\Ensure $S_{best}$
\vspace{2mm}
\Function{DE}{$\mathit{np}$, $f$, $cr$, $\mathit{life}$, $\mathit{Goal}$}
 \State $Population  \gets $ InitializePopulation($\mathit{np}$)   
 \State $S_{best} \gets $GetBestSolution($Population $)
 \While{$\mathit{life} > 0$}
\State $NewGeneration \gets \emptyset$
\For{$i=0 \to \mathit{np}-1$}
\State $S_i \gets$ Extrapolate($Population [i], Population , cr, f$)
\If {Score($S_i$) >Score($Population [i]$)}
\State $NewGeneration$.append($S_i$)
\Else
\State $NewGeneration$.append($Population [i]$)
\EndIf
\EndFor
\State $Population  \gets NewGeneration$
\If{$\neg$ Improve($Population $)}
\State $life -=1$
\EndIf
\State $S_{best} \gets$ GetBestSolution($Population $)
 \EndWhile
\State \Return $S_{best}$
\EndFunction
\Function{Score}{$Candidate$}
   \State set tuned parameters according to $Candidate$
   \State $model \gets$TrainLearner()
   \State $result \gets$TestLearner($model$)   
   \State \Return$\mathit{Goal}(result)$  
\EndFunction
\Function{Extrapolate}{$old, pop, cr, f$}
  \State $a, b, c\gets threeOthers(pop,old)$  
  \State $newf \gets \emptyset$
  \For{$i=0 \to \mathit{np}-1$}
       \If{$cr < random()$}
         \State $newf$.append($old[i]$)
                \Else
                  \If{typeof($old[i]$) == bool}
                    \State $newf$.append(not $old[i]$)
         \Else
          \State $newf$.append(trim($i$,($a[i] + f * (b[i] - c[i]$)))) 
         \EndIf
       \EndIf
  \EndFor
 \State \Return $newf$
\EndFunction
        \end{algorithmic} 
\caption{Pseudocode for DE with Early Termination}
\label{alg:DE}
\end{algorithm}

%  \begin{table*}[!t]

% \renewcommand{\baselinestretch}{0.8}
% \scriptsize
% \centering
%   \begin{tabular}{c c c c c c c c c c }\hline
%   Dataset &antV0&antV1&antV2&camelV0&camelV1&ivy&jeditV0&jeditV1&jeditV2
% \\\hline
%   training &20/125 &40/178 &32/293 &13/339 &216/608 &63/111 &90/272 &75/306 &79/312
% \\  tuning  &40/178 &32/293 &92/351 &216/608 &145/872 &16/241 &75/306 &79/312 &48/367
% \\  testing &32/293 &92/351 &166/745 &145/872 &188/965 &40/352 &79/312 &48/367 &11/492
% \\ \hline
%   Dataset &log4j&lucene&poiV0&poiV1&synapse&velocity&xercesV0&xercesV1
% \\\hline
%   training &34/135 &91/195 &141/237 &37/314 &16/157 &147/196 &77/162 &71/440
% \\  tuning  &37/109 &144/247 &37/314 &248/385 &60/222 &142/214 &71/440 &69/453
% \\  testing &189/205 &203/340 &248/385 &281/442 &86/256 &78/229 &69/453 &437/588
% \\  \end{tabular}

%   \caption{Data used in this experiment. 
%   E.g., the top left data set has 20 defective classes out of 125 total.
%   See \tion{dataa} for explanation of {\em training, tuning, testing} sets.
%   }\label{tab:data1}
% \end{table*} 

 \begin{table*}[!t]

\renewcommand{\baselinestretch}{0.8}
\scriptsize
\centering
%   \begin{tabular}{p{0.75cm}p{0.75cm}p{0.75cm}p{0.75cm}p{0.75cm}p{0.75cm}p{0.75cm}p{0.75cm}p{0.75cm}p{0.75cm}}\hline
  \begin{tabular}{c c c c c c c c c c } \hline
  Dataset &antV0&antV1&antV2&camelV0&camelV1&ivy&jeditV0&jeditV1&jeditV2
\\\hline
  training &20/125 &40/178 &32/293 &13/339 &216/608 &63/111 &90/272 &75/306 &79/312
\\  tuning  &40/178 &32/293 &92/351 &216/608 &145/872 &16/241 &75/306 &79/312 &48/367
\\  testing &32/293 &92/351 &166/745 &145/872 &188/965 &40/352 &79/312 &48/367 &11/492
\\ \hline
  Dataset &log4j&lucene&poiV0&poiV1&synapse&velocity&xercesV0&xercesV1
\\\hline
  training &34/135 &91/195 &141/237 &37/314 &16/157 &147/196 &77/162 &71/440
\\  tuning  &37/109 &144/247 &37/314 &248/385 &60/222 &142/214 &71/440 &69/453
\\  testing &189/205 &203/340 &248/385 &281/442 &86/256 &78/229 &69/453 &437/588
\\  \end{tabular}

   \caption{Data used in this experiment. 
   E.g., the top left data set has 20 defective classes out of 125 total.
   See \tion{dataa} for explanation of {\em training, tuning, testing} sets.
   }\label{tab:data1}
\end{table*} 

DE evolves a {\em NewGeneration} of candidates  from
a current {\em Population}.  Our DE's lose one ``life''
when the new population is no better than  current one (terminating when ``life'' is zero)$^{L4}$.
Each candidate solution in the {\em Population}  
is a pair of {\em (Tunings, Scores)}.  {\em Tunings} are selected from
\tab{parameters} and {\em Scores} come from training a learner using those parameters
and applying it     test data$^{L23-L27}$.

The premise of DE  is that the best way to mutate the existing tunings
is to {\em Extrapolate}$^{L28}$
between current solutions.  Three solutions $a,b,c$ are selected at random.
For each tuning parameter $i$, at some probability {\em cr}, we replace
the old tuning $x_i$ with $y_i$. For booleans, we use $y_i= \neg x_i$ (see line 36). For numerics, $y_i = a_i+f \times (b_i - c_i)$   where $f$ is a parameter
controlling  cross-over.  The {\em trim} function$^{L38}$ limits the new
value to the legal range min..max of that parameter.
 
The main loop of DE$^{L6}$ runs over the {\em Population}, replacing old items
with new {\em Candidate}s (if  new candidate is better).
This means that, as the loop progresses, the {\em Population} is full of increasingly
more valuable solutions. This, in turn, also improves  the candidates, which are {\em Extrapolate}d
from the {\em Population}.

For the experiments of this paper, we collect performance
values from a data mining, from which a {\em Goal} function extracts one 
performance value$^{L26}$ (so we run this code many times, each time with
a different {\em Goal}$^{L1}$).  Technically, this makes a  {\em single objective} DE (and for notes on multi-objective DEs, see~\cite{Coello05,zhang07,5583335}).

%\begin{algorithm}
%\begin{algorithmic}[1]
% \KwData{this text}
% \KwResult{how to write algorithm with \LaTeX2e }
% initialization\;
% \While{not at end of this document}{
%  read current\;
%  \eIf{understand}{
%   go to next section\;
%   current section becomes this one\;
%   }{
%   go back to the beginning of current section\;
%  }
% }
% \caption{How to write algorithms}
% \end{algorithmic}
%\end{algorithm}

\section{Experimental Design}\label{sect:design}

\subsection{Data Sets}\label{sect:dataa}

Our defect data comes from the PROMISE repository (http://openscience.us/repo/defect)
and pertains to 
open source Java systems defined in terms of \tab{ck}:  {\it ant}, {\it camel}, {\it ivy}, {\it jedit}, {\it log4j}, {\it lucene},
{\it poi}, {\it synapse}, {\it velocity} and {\it xerces}. 

An important principle in data mining is not to test on the data used
in training.  There are many ways to design a experiment that satisfies this principle.
Some of those methods have  limitations; e.g.
{\em leave-one-out} is too slow for large data sets and
{\em cross-validation} mixes up older and newer data  (such that
data from the {\em past} may be used to test on {\em future data}).

To avoid these problems, we used an incremental learning approach. The following
experiment ensures that the training data was created at some time before the test
data.
For this experiment, we use data sets with at least three  
consecutive releases  (where release $i+1$ was built after release $i$). When tuning a learner,
\bi 
% \item {\em Tuned learner}: the {\em first} release is used  on line 24 of Algorithm~\ref{alg:DE} to
%   build some model using some the tunings found in some {\em Candidate}. The {\em second} release was used on line 25 of Algorithm~\ref{alg:DE} to 
%   test the model found on line 24. The {\em third} release was used as a testing data to gather the performance statistics
%   reported below from the best model found by DE.
% \item {\em untuned learner}:

\item The {\em first} release was used  on line 24 of Algorithm~\ref{alg:DE} to
   build some model using some the tunings found in some {\em Candidate}.
 \item The {\em second} release was used on line 25 of Algorithm~\ref{alg:DE} to 
   test the candidate model found on line 24.
   \item Finally the {\em third} release was used to gather the performance statistics
   reported below from the best model found by DE.
 \ei
 
 To be fair for the untuned learner, the {\em first} and {\em second} releases used in tuning experiments
 will be combined as the training data to build a model. Then the performance of this untuned learner 
 will be evaluated by the same {\em third} release as in the tuning experiment.
 
% This approach ensures   all treatments 
% are assessed on the same tests. Note that we did consider
% one other experimental design but rejected it for reasons of
% internal validity (see \tion{construct}).

Some data sets have more than three releases and, for those data, we could run more
 than one experiment. For example, {\em ant} has five versions in PROMISE so
 we ran three experiments called V0,V1,V2:
 \bi
 \item AntV0: first, second, third = versions 1, 2, 3
 \item AntV1: first, second, third = versions 2, 3, 4
 \item AntV2: first, second, third = versions 3, 4, 5
 \ei 
These data sets are displayed in \tab{data1}.

% As an aside, an alternate experimental design would be to 
% learn a baseline learner from the first {\em and} second release
% instead of, as shown above,  just the first release. On the one hand,
% this would mean that the baseline could be learned from more data.
% On the other hand, this adds a conflation to our experimental design
% since the optimizer uses the second release for pruning, not growing a data set.  Happily, from 
% piror work~\cite{Menzies:2008aa} we know that defect predictors usually {\em saturates} (i.e.
% does not generate better predictors) after 100 examples, which is a number smaller than all our first release data sets. Hence, their would
% be little value in generating the baselines using the first and second
% releases. 

\subsection{Optimization Goals}

Recall from Algorithm~1 that we call differential evolution once for each
 optimization goal. This section lists those optimization goals.
Let $\{A,B,C,D\}$ denote the
true negatives, 
false negatives, 
false positives, and 
true positives
(respectively) found by a binary detector. 
Certain standard measures can be computed from
$A,B,C,D$, as shown below. Note that for $pf$, the {\em better} scores are {\em smaller}
while
for all other scores, the {\em better} scores are {\em larger}.

{\scriptsize\[
\begin{array}{ll}
pd=recall=&D/(B+D)\\
pf=&C/(A+C)\\ 
prec=precision=&D/(D+C) \\
F =&2*pd*prec/(pd + prec)
\end{array}
\]}

The rest of this paper explores tuning for {\em prec} and {\em F}. As discussed
in \tion{goals}, our point is not that these are best or most important optimization goals.
Indeed, the list of ``most important'' goals is domain-specific (see \tion{goals})
and we only explore these two to illustrate how conclusions can change dramatically
when moving from one goal to another.

\section{Experimental Results}

In the following, we explore the effects of tuning WHERE, Random Forest,
and CART. LR will be used, untuned, in order to check one of the recommendations
made by Hall et al.~\cite{hall11}.

\subsection{RQ1:  Does  Tuning  Improve Performance? }\label{sect:precision}

\begin{figure}[!t]
\begin{center}
\includegraphics[width=1.5in]{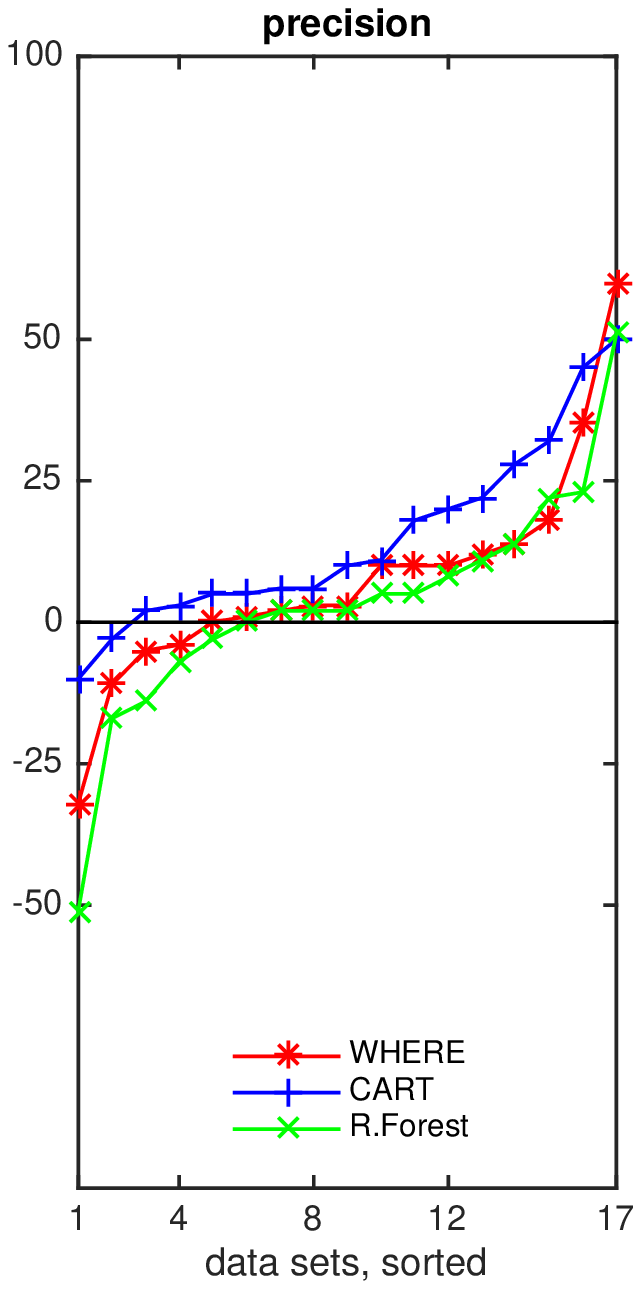}\includegraphics[width=1.5in]{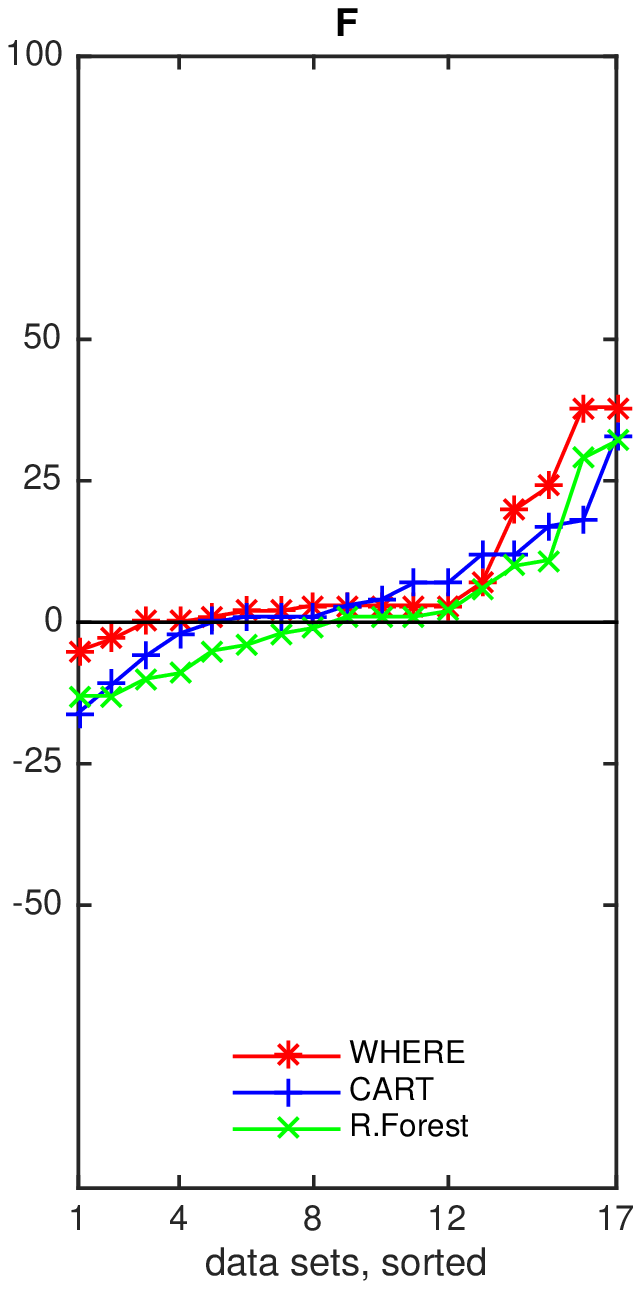}
 \end{center}
\caption{Deltas in performance  seen in \tab{precisionbars} (left)
and \tab{fbars} (right) between tuned and untuned learners. Tuning improves performance when the deltas are above zero.}\label{fig:deltas}
 \end{figure}

\fig{deltas} says  that the answer to RQ1 is ``yes''-- tuning  has a positive effect on performance scores. This figure sorts
 deltas in the precision and the F-measure    between tuned and untuned learners. Our reading of this
figure is that, overall, tuning rarely makes performance   worse and often can make it much better.

\tab{precisionbars} and \tab{fbars} show the
the specific values seen before and after tuning with {\em precision}
and {\em ``F''} as different optimization goals(the corresponding  ``F'' and precision values for
\tab{precisionbars} and \tab{fbars} are not provided for the space limitation).
For each data set, the maximum precision or ``F'' values for each data set are shown in {\bf bold}.
As might have been
 predicted by Lessmann et al.~\cite{lessmann2008benchmarking}, 
untuned CART is indeed the worst learner (only one of its
untuned results is best and {\bf bold}). 
And, 
in $\frac{12}{17}$ cases, the  untuned Random Forest performs better than or equal to untuned CART in terms of precision.  
% Note that the size of the improvement is sometimes small; e.g. the precision results improve more than the ``F'' measure.
% But even when the {\em median} change is small, there still may exist
% interesting (i.e.  exceptionally
% large) improvements from tuning-- for example, 
%  see the last  two ``F'' improvements for WHERE in \fig{deltas} where the improvements were greater than 70\%.
% Some of the data sets in \fig{precisionbars} proved challenging for all learners;
% e.g. the precision results for {\em ivy} are less
% that impressive.
% To some extent, this is due to
% the properties of   the data set (as shown in \fig{data1}, defective classes in {\em ivy} are very rare in tuning data).

That said,  tuning can improve those poor performing detectors.
In some cases, the median changes may be small (e.g. the ``F'' results for WHERE and Random Forests) but even in
those cases, there are enough large changes to motivate the use of tuning. For example:
\bi
\item
For ``F'' improvement, there are two improvements over 25\% for both WHERE and Random Forests. Also, in {\em poiV0}, all untuned learners report ``F'' of under 50\%, tuning changes those scores by 25\%. Finally, note the  {\em xercesV1} result for the WHERE learner. Here, tuning changes precision from 32\% to 70\%.
\item
Regarding precision, for {\em antV0}, and {\em antV1} untuned WHERE reports precision of 0. But tuned WHERE scores 35 and 60 (the similar pattern can seen in ``F'').

\ei

% To complete the discussion in this section, we note that in the Lessmann et al. study,  Random Forest dramatically out-performed CART.
% In this study, we show that  
% tuned CART is now comparable to Random Forest. So our new results
% do not complete reverse the results of Lessmann et al. However, they
% do show that the Lessmann results are ``brittle'', in the sense that
% tuning can remove the effect they report.

\begin{table}[!t]
\renewcommand{\baselinestretch}{0.8} 
% \centering
\scriptsize    

\begin{tabular}{r|rl|rl|rl|rl|rl|rlrl}
%\begin{tabular}{r@{~}|r@{~}l@{~}|r@{~}l@{~}|r@{~}l|r@{~}l@{~}|r@{~}l@{~}|r@{~}l@{~}r@{~}l}
      &   \multicolumn{4}{c|}{WHERE}         &   \multicolumn{4}{c|}{CART}         &   \multicolumn{4}{c}{Random Forest}         \\\hline
  Data set   &   \multicolumn{2}{c}{default}         &   \multicolumn{2}{c|}{Tuned}         &   \multicolumn{2}{c}{default}         &   \multicolumn{2}{c|}{Tuned}    &   \multicolumn{2}{c}{default}  &   \multicolumn{2}{c}{Tuned}\\\hline
antV0 & 0 &   & 35 &   & 15 &   & {\bf 60} &   & 21 &   & 44 &  \\
antV1 & 0 &   & 60 &   & 54 &   & 56 &   & {\bf 67} &   & 50 &  \\
antV2 & 45 &   & 55 &   & 42 &   & 52 &   & 56 &   & {\bf 67} &  \\
camelV0 & 20 &   & 30 &   & 30 &   & 50 &   & 28 &   & {\bf 79} &  \\
camelV1 & 27 &   & 28 &   & {\bf 38} &   & 28 &   & 34 &   & 27 &  \\
ivy & 25 &   & 21 &   & 21 &   & {\bf 26} &   & 23 &   & 20 &  \\
jeditV0 & 34 &   & 37 &   & 56 &   & {\bf 78} &   & 52 &   & 60 &  \\
jeditV1 & 30 &   & 42 &   & 32 &   & {\bf 64} &   & 32 &   & 37 &  \\
jeditV2 & 4 &   & {\bf 22} &   & 6 &   & 17 &   & 4 &   & 6 &  \\
log4j & 96 &   & 91 &   & 95 &   & 98 &   & 95 &   & {\bf 100} &  \\
lucene & 61 &   & 75 &   & 67 &   & 70 &   & 63 &   & {\bf 77} &  \\
poiV0 & 70 &   & 70 &   & 65 &   & {\bf 71} &   &  67 &   & 69 &  \\
poiV1 & 74 &   & 76 &   & 72 &   & 90 &   & 78 &   & {\bf 100} &  \\
synapse & 61 &   & 50 &   & 50 &   & {\bf 100} &   & 60 &   & 60 &  \\
velocity & 34 &   & {\bf 44} &   & 39 &   & {\bf 44} &   & 40 &   & 42 &  \\
xercesV0 & 14 &   & 17 &   & 17 &   & 14 &   & {\bf 28} &   & 14 &  \\
xercesV1 & 86 &   & 54 &   & 72 &   & {\bf 100} &   & 78 &   & 27 &  \\
\end{tabular}
\caption{Precision results (best results  shown in {\bf bold}).}
\label{tab:precisionbars}
\end{table}

\begin{table}[!t]
\renewcommand{\baselinestretch}{0.8} 
% \centering
\scriptsize  
~~~\begin{tabular}{r|rl|rl|rl|rl|rl|rlrl}
      &   \multicolumn{4}{c|}{WHERE}         &   \multicolumn{4}{c|}{CART}         &   \multicolumn{4}{c}{Random Forest}         \\\hline
  Data set   &   \multicolumn{2}{c}{default}         &   \multicolumn{2}{c|}{Tuned}         &   \multicolumn{2}{c}{default}         &   \multicolumn{2}{c|}{Tuned}    &   \multicolumn{2}{c}{default}  &   \multicolumn{2}{c}{Tuned}\\\hline
antV0 & 0 &   & 20 &   & 20 &   & {\bf 40} &   & 28 &   & 38 &  \\
antV1 & 0 &   & 38 &   & 37 &   & {\bf 49} &   & 38 &   & {\bf 49} &  \\
antV2 & 47 &   & 50 &   & 45 &   &  49 &   & {\bf 57} &   & 56 &  \\
camelV0 & 31 &   & 28 &   & 39 &   & 28 &   & {\bf 40} &   & 30 &  \\
camelV1 & 34 &   & 34 &   & 38 &   & 32 &   & {\bf 42} &   & 33 &  \\
ivy & 39 &   & 34 &   & 28 &   & {\bf 40} &   & 35 &   &  33 &  \\
jeditV0 & 45 &   & 47 &   & 56 &   & 57 &   & {\bf 63} &   & 59 &  \\
jeditV1 & 43 &   & 44 &   & 44 &   & 47 &   & 46 &   & {\bf 48} &  \\
jeditV2 & 8 &   & {\bf 11} &   & 10 &   & 10 &   & 8 &   & 9 &  \\
log4j & 47 &   & 50 &   & 53 &   & 37 &   & {\bf 60} &   & 47 &  \\
lucene & 73 &   & 73 &   & 65 &   & 72 &   & 70 &   & {\bf 76} &  \\
poiV0 & 50 &   & 74 &   & 31 &   & 64 &   & 45 &   & {\bf 77} &  \\
poiV1 & 75 &   & {\bf 78} &   & 68 &   & 69 &   & 77 &   & {\bf 78} &  \\
synapse & 49 &   & 56 &   & 43 &   & {\bf 60} &   & 52 &   & 53 &  \\
velocity & 51 &   & 53 &   & 53 &   & 51 &   & {\bf 56} &   & 51 &  \\
xercesV0 & 19 &   & 22 &   & 19 &   & {\bf 26} &   & 34 &   & 21 &  \\
xercesV1 & 32 &   & 70 &   & 34 &   & 35 &   & 42 &   & {\bf 71} &  \\
\end{tabular}
\caption{F-measure results (best results  shown in {\bf bold}).}
\label{tab:fbars}
\end{table}

\subsection{RQ2:  Does Tuning Change a Learner's Ranking ?}\label{sect:rank}
Researchers often use performance criteria to assert that one learner is better than 
another~\cite{lessmann2008benchmarking,me07b,hall11}. For example:
\be
\item
Lessmann et al.~\cite{lessmann2008benchmarking} conclude that
Random Forest is considered to be statistically 
better than CART. 
\item
Also, in Hall et al.'s   systematic literature review\cite{hall11}, it is argued
that defect predictors based on simple 
modeling techniques such as LR perform better than ``complicated'' techniques such as Random Forest.
To explain that comment, we note that
by three measures,
Random Forest
is more complicated than LR:
\be
\item
CART builds one model
while Random Forest builds many models. 
\item
LR is just
a model construction tool while Random Forest needs both
a tool to construct its forest {\em and} a second tool
to  infer some conclusion from all the members of that forest.
\item
the LR model can be printed in a few lines while the multiple
models learned by Random
Forest model would take up multiple pages of output.
\ee
\ee
Given tuning, how stable are these  conclusions?
Before answering this issue, we digress for two comments.

Firstly, it is important to comment on why it is  so important to check the conclusions
of these particular papers. 
These  papers are prominent publications (to say the least).
Hall et al.~\cite{hall11} is the fourth most-cited IEEE TSE
paper for 2009 to 2014 with 176 citations (see goo.gl/MGrGr7)
while the Lessmann et al. paper~\cite{lessmann2008benchmarking} has 394 citations (see
goo.gl/khTp97)-- which is quite remarkable for a paper published in 2009.
Given the prominence
of these papers, researchers might believe it is
appropriate to
  use  their advice without testing that advice on local data sets.

Secondly, while we are critical of the results of
Lessmann et al. and Hall et al., it needs to be said that  their analysis  was 
excellent and exemplary given the state-of-the-art of the tools used when those papers were written.  
While Hall et al. did not perform any new experiments, 
their
summarization of so many defect prediction papers has not been equalled
before (or since).
As to the Lessmann et al. paper, they  compared
22 data miners using various   data sets (mostly from NASA)~\cite{lessmann2008benchmarking}.
In that study, some learners were tuned using manual methods 
(C4.5, CART and Random Forest)
and some, like SVM-Type learners, were tuned by automatic grid search (for more on grid search, see   {\S}2.1).

\begin{figure}[!t]
% \begin{center}
\includegraphics[width=1.5in]{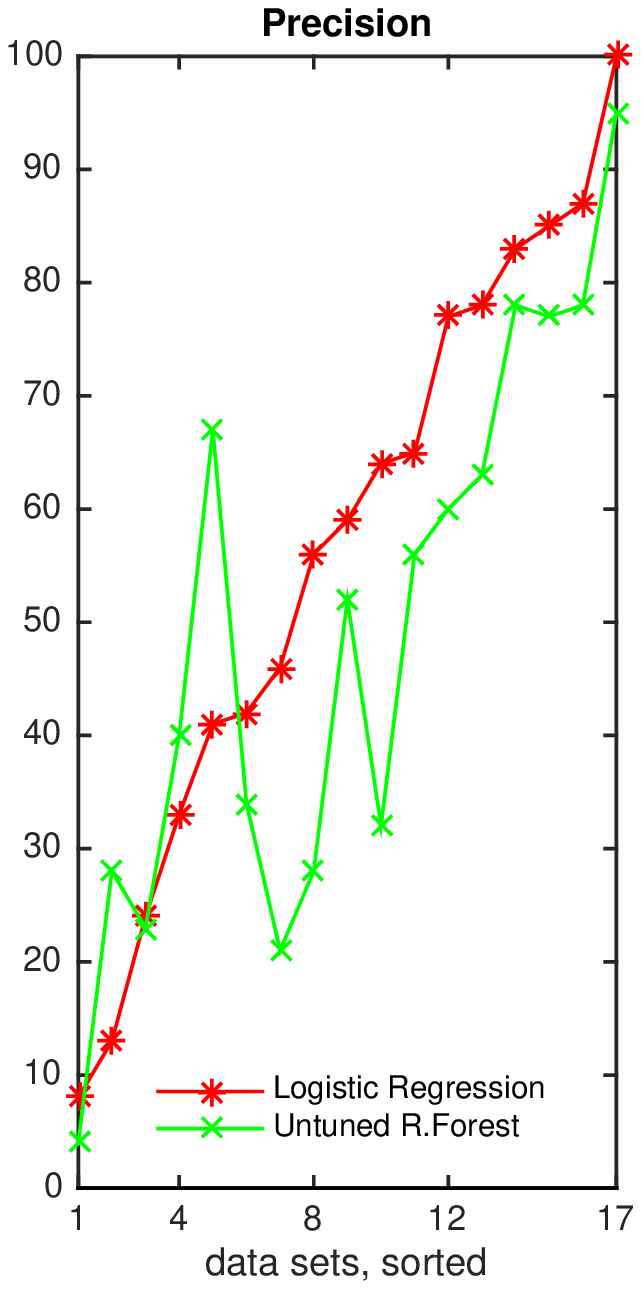}\includegraphics[width=1.5in]{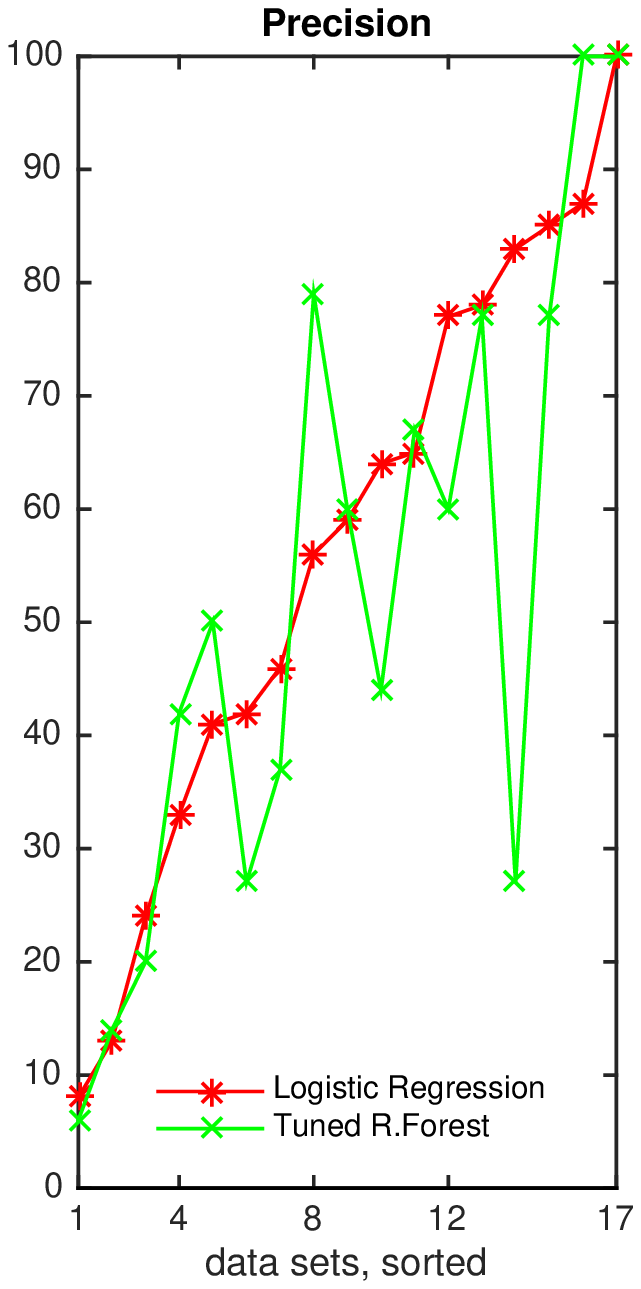}
%  \end{center}
\caption{Comparison between Logistic Regression and Random Forest before and after tuning. }\label{fig:lr}
 \end{figure}

That said, our tuning results show that it is time to revise
the recommendations of those papers. 
  \fig{lr} comments on the advice from Hall et al. (that LR is better than Random Forest)L
  \bi
  \item
 In a result that might have been predicted by Hall et al.,  
 untuned Random
Forests performs comparatively worse than
 Logistic Regression. Specifically, untuned
 Random Forest performs worse than Logistic regression in  13 out of   17 data sets. 
\item
However, it turns out that advice is sensitive to the tunings
used with Random Forest. After tuning, we find that tuned Random Forest
loses to Logistic Regression in only 6 out of 17 data sets. 
\ei

As to Lessmann et al.'s advice (that Random Forest is better than 
CART),  
in \tab{precisionbars} and \tab{fbars}, we saw those counter-examples
to that statement.
Recall in those tables,
tuned CART are better than or equal
to tuned Random Forest in $\frac{12}{17}$ and $\frac{7}{17}$ data sets in
terms of precision and F-measure, respectively. Prior to tuning experiments, those
numbers are $\frac{5}{17}$ and $\frac{1}{17}$. Results from the non-parametric
Kolmogorov-Smirnov(KS) Test show that the performance 
scores of tuned CART and tuned Random Forest are not statistically different.
Note that Random Forest is  not significantly better than CART, which would not have been
predicted by   Lessmann et al.

Hence we answer RQ2 as ``yes'': tuning can change how  data miners are comparatively ranked.

% \begin{figure}[!t]
% \begin{center}
% \includegraphics[width=1.5in]{svm_rbf.eps}\includegraphics[width=1.5in]{svm_sigmoid.eps}
%  \end{center}
% \caption{Comparison between tuned and naive SVM learners with rbf and sigmoid kernels over the goal of F. }\label{fig:svm}
%  \end{figure}

 \subsection{RQ3: Does Tuning Select Different Project Factors? }\label{sect:import}

Researchers often use data miners to  test what factors have most impact on software projects~\cite{bell2013limited,rahman2013how,me02k,Moser:2008,zimmermann2007predicting,herzig2013predicting}. 
\tab{features} comments that such tests are unreliable since the factors selected by a data miner are much altered before and 
after tuning.

\tab{features} shows what features are found in the trees generated by the WHERE algorithm
(bold shows the features found by the trees from tuned WHERE; plain text shows the features seen
in the untuned study). Note that different features are selected depending on whether or not
we tune an algorithm.

% list some features
\begin{table}[!t]

\renewcommand{\baselinestretch}{0.8}
\scriptsize
\centering
  \begin{tabular}{c|p{1in}|p{1in}}
    \multicolumn{1}{c|}{ Data set}  &   \multicolumn{1}{c|}{Precision} & \multicolumn{1}{c}{F} \\ \hline 
 \multirow{2}{*}{antV0} & {\bf rfc} &  {\bf None} \\
         & mfa, loc, cam, dit, dam, lcom3 & mfa, loc, cam, dit, dam, lcom3\\
  \hline
 \multirow{2}{*}{camelV0} & {\bf mfa, wmc, lcom3} &{\bf None }\\
        & mfa, wmc, rfc, loc, cam, lcom3 & mfa, wmc, rfc, loc, cam, lcom3\\
  \hline
 \multirow{2}{*}{ivy} & {\bf cam, dam, npm, loc, rfc, wmc} &{\bf cam, dam, npm, loc, rfc, wmc }  \\
       & loc, cam, dam, wmc, lcom3 & loc, cam, dam, wmc, lcom3 \\
  \hline
 \multirow{2}{*}{jeditV0} &{\bf mfa, dam, loc }&{\bf mfa, dam, loc}\\
         & mfa, lcom3, dam, dit, ic & mfa, lcom3, dam, dit, ic \\
  \hline
 \multirow{2}{*}{log4j} & {\bf loc, ic, dit }&{\bf mfa, wmc, rfc, loc, npm}\\
         & mfa, lcom3, loc, ic & mfa, lcom3, loc, ic \\
   \hline
  \multirow{2}{*}{lucene} & {\bf dit, cam, wmc, lcom3, dam, rfc, cbm, mfa, ic} & {\bf dit, lcom3, dam, mfa}\\
         & dit, cam, dam, ic & dit, cam, dam, cbm, ic\\
   \hline
   \multirow{2}{*}{poiV0} & {\bf mfa, amc, dam }& {\bf mfa, amc, dam} \\
        & mfa, loc, amc, dam, wmc, lcom & mfa, loc, amc, dam, wmc, lcom\\
   \hline
   \multirow{2}{*}{synapse} &{\bf loc, dit, rfc, cam, wmc, dam, lcom,  mfa,  lcom3} & {\bf dam} \\
        & loc, mfa, cam, lcom, dam, lcom3 & loc, mfa, cam, lcom, dam, lcom3\\
    \hline
   \multirow{2}{*}{velocity} & {\bf dit, wmc, cam, rfc, cbo, moa, dam }& {\bf mfa, dit }\\
        & dit, dam, lcom3, ic, mfa, cbm  & dit, dam, lcom3, ic, mfa\\
    \hline
   \multirow{2}{*}{xercesV0} &{\bf  wmc }&{\bf cam, dam, avg\_cc, loc, wmc,  dit,  mfa, ce, lcom3 }\\
        & wmc, mfa, lcom3, cam, dam &  wmc, mfa, lcom3, cam, dam \\
    \hline
    
  \end{tabular}
  
    \caption{Features selected by tuned WHERE with different goals:
    {\bf bold} features are those found useful by the tuned WHERE.
    Also, features shown in plain text are those found useful by the untuned WHERE.
    }\label{tab:features}
\end{table}

For example, consider {\em mfa} which is the
number of methods inherited by a class plus the number of methods accessible by member methods of the class.
For both goals (precision and ``F'') {\em mfa} is selected for 8 and 5 data sets,
for the untuned and tuned data miner (respectively).
Similar differences are   seen with other attributes.

As to why different tunings select for different features,  recall from {\S}2.1 that tuning changes how data miners
heuristically explore a large space of possible models. As we change how that exploration proceeds,
so we change what features are found by that exploration.

In any case, our answer to RQ3 is ``yes'', tuning changes our
conclusions about what factors are most important in software engineering.
Hence, many old papers    need to be revisited  and perhaps revised~\cite{bell2013limited,rahman2013how,me02k,Moser:2008,zimmermann2007predicting,herzig2013predicting}.  
For example, one of us (Menzies) used data miners
to assert that some factors were more important than others for predicting
successful software reuse~\cite{me02k}. That assertion should now be doubted since Menzies did not conduct a tuning study before reporting what factors the data miners
found were most influential.

\begin{figure}[!t]
% \begin{center}
\includegraphics[width=1.5in]{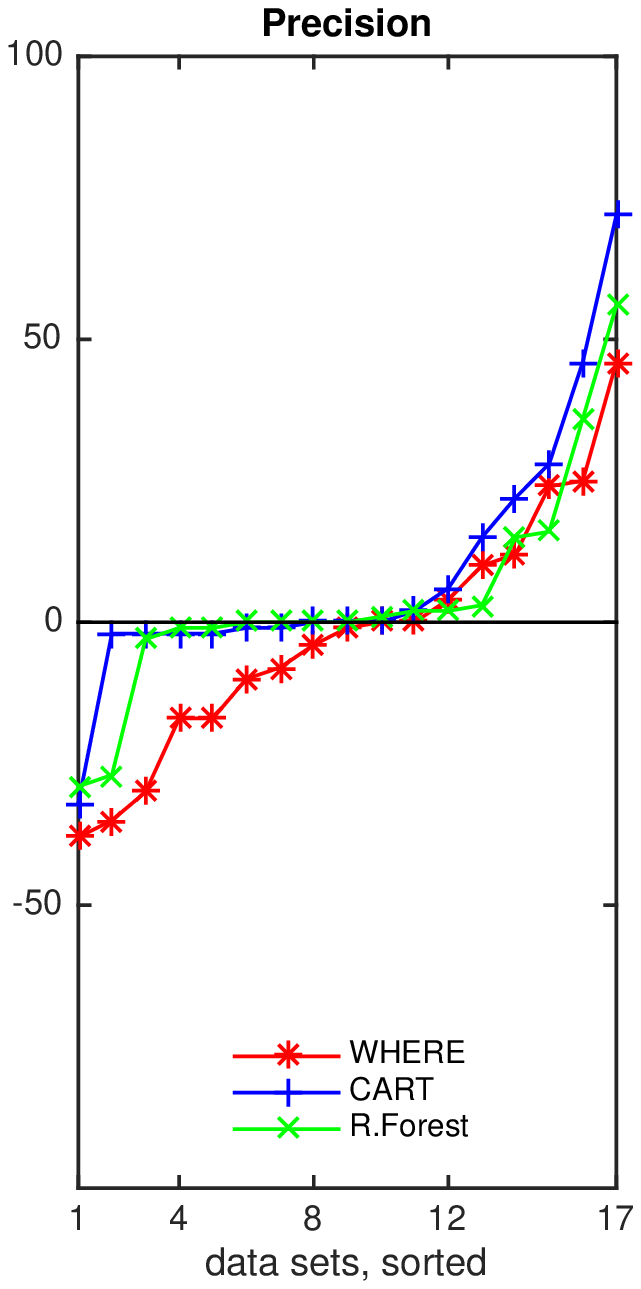}\includegraphics[width=1.5in]{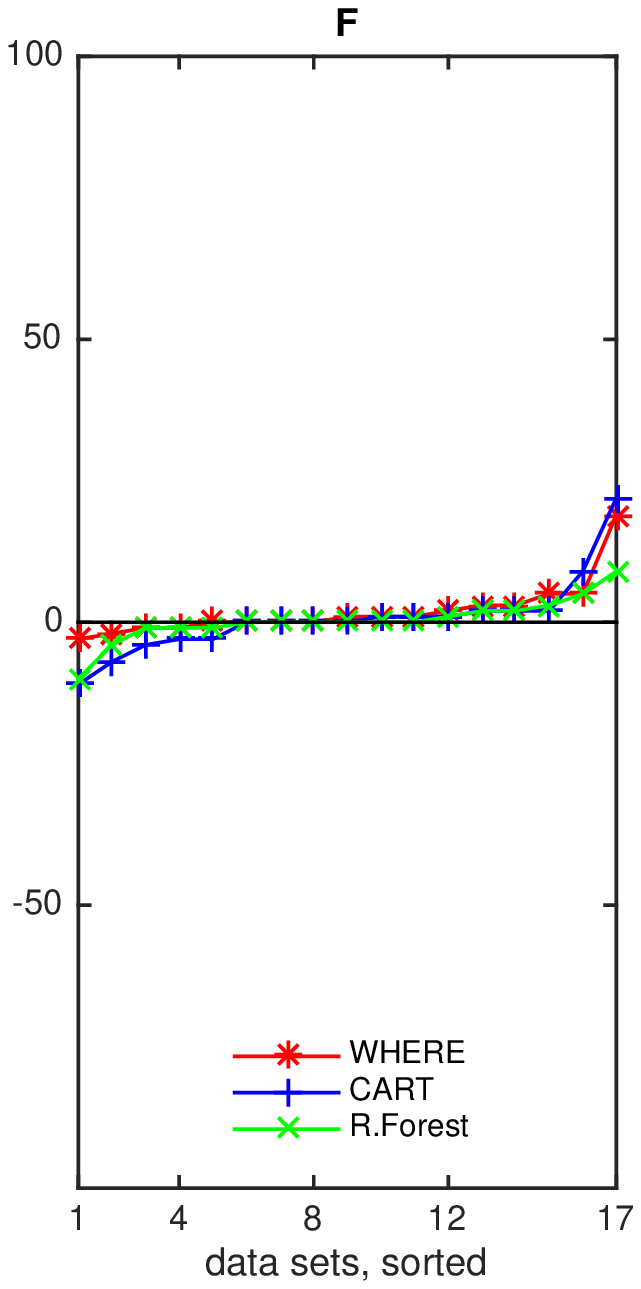}
%  \end{center}
\caption{Deltas in performance between {\em np = 10} and the recommended np's. The recommended np is better when deltas are above zero. {\em np = 90, 50 and 60} are recommended population size for WHERE, CART and Random Forest by Storn.}\label{fig:deltas_np}
 \end{figure}

\subsection{RQ4: Is Tuning Easy?}\label{sect:easy}

In terms of the search space
explored via tuning, optimizing defect prediction from static code
measures is much {\em smaller} than the standard optimization.

To see this,
recall from Algorithm~1 that
DE explores a {\em Population} of size {\em np = 10}. This is a very small population size since
Rainer Storn (one of the inventors of DE) recommends  setting {\em np} to be ten times larger than the number
of attributes being optimized~\cite{storn1997differential}.

From \tab{parameters},
we see that Storn would therefore recommend {\em np} values of
90, 50, 60 for WHERE, CART and Random Forest (respectively). Yet we achieve our results
using a constant {\em np = 10}; i.e. $\frac{10}{90}, \frac{10}{50}, \frac{10}{60}$ of the
recommended search space.

To justify that {\em np = 10} is enough, we did another tuning study, 
where all the settings were the same as before but we set {\em np = 90, np = 50} and {\em np = 60} for WHERE, CART and Random Forest, respectively (i.e. the settings
as recommended by Storn). The tuning performance of learners was evaluated
by precision and ``F'' as before. To compare performance of each learner with different {\em np}'s, we computed the delta in the performance between {\em np = 10} and {\em np} using any of \{90, 50, 60\}.

Those deltas, shown in \fig{deltas_np}, are sorted along the x-axis. In those plots, a zero or negative $y$ value means that {\em np = 10} performs as well or better than  {\em np $\in \{90, 50, 60\}$}. One technical aside: the data set orderings in \fig{deltas_np} on the x-axis are not the same (that is,
if {\em np $>$ 10} was useful for optimizing one data set's precision score, it was not necessary for that data set's F-measure score).

  \fig{deltas_np} shows that
the median improvement is zero; i.e. {\em np = 10} usually does as well as anything else. This observation is
supported by the   KS
  results of \tab{ks}. At a 95\% confidence, the
KS threshold  is $1.36\sqrt{34/(17*17)} = 0.46$, which is greater than  the values in \fig{deltas_np}. That is, no result in  \fig{deltas_np} is significantly different to any other-- which is to say that
there is no evidence that   {\em np = 10} is a poor choice of search space size.

Another measure showing that tuning is easy 
(for static code defect predictors)
is the number of evaluations required to complete optimization
(see next section). That is, we answer RQ4 as ``yes'', tuning is surprisingly easy-- at least
for defect predictors and using DE.

% \begin{table}[!t]
% \centering
% \scriptsize
% \begin{tabular}{|l|c|c|}
% \hline
% Learner & CART & WHERE \\ \hline
% CART    & -    & 0.42  \\ \hline
% R.Forest      & 0.29 & 0.18  \\ \hline
% \end{tabular}
% \begin{tabular}{|l|c|c|}
% \hline
% Learner & CART & WHERE \\ \hline
% CART    & -    & 0.24  \\ \hline
% R.Forest      & 0.24 & 0.29  \\ \hline
% \end{tabular}
% \caption{Kolmogorov-Smirnov Tests for numbers seen in \fig{deltas_np}: Precision(left) and F(right).}
% \end{table}

\begin{table}[!t]
\renewcommand{\baselinestretch}{0.8}
\scriptsize
 \centering
  \begin{tabular}{c|c c|cc}
    &   \multicolumn{2}{c|}{Precision} & \multicolumn{2}{c}{F} \\ \hline 
    Learner & CART  & WHERE & CART & WHERE \\
\hline
    CART & - & 0.41 & - & 0.24 \\
    R. Forest &  0.12 & 0.35 & 0.18 & 0.18 \\
  \end{tabular}
    \caption{Kolmogorov-Smirnov Tests for distributions of  \fig{deltas_np}}\label{tab:ks}
\end{table}

\begin{figure}[!t]
\begin{center}
\includegraphics[width=1.5in]{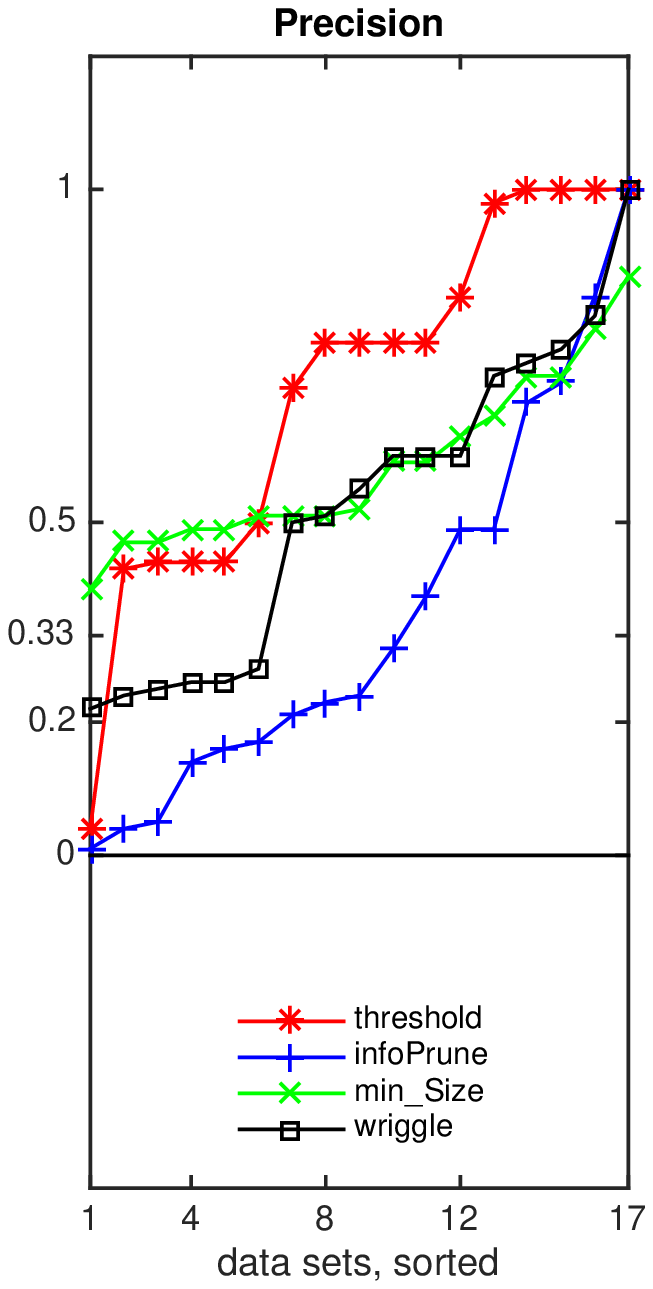}\includegraphics[width=1.5in]{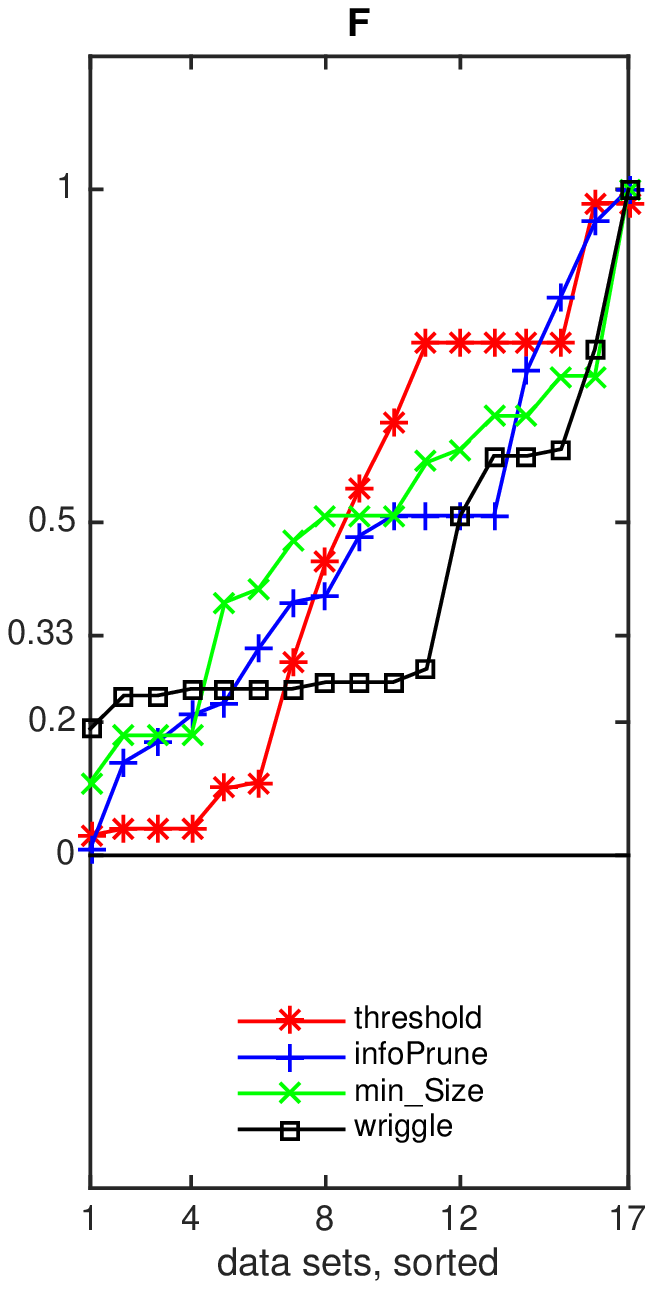}
\end{center}
\caption{Four representative tuning values in WHERE with  precision and F-measure as the tuning goal, respectively.   }\label{fig:features}
 \end{figure}

\subsection{RQ5: Is Tuning Impractically Slow?}\label{sect:fast}

%%%%evaluations for pre and F %%%%%%
\begin{table}[!ht]
\renewcommand{\baselinestretch}{0.75}
\scriptsize
\centering
  \begin{tabular}{r|c c|c c|c c }
    Datasets & \multicolumn{2}{c|}{Tuned\_Where}  & \multicolumn{2}{c|}{Tuned\_CART} &  \multicolumn{2}{c}{Tuned\_RanFst}  \\
    \hline
    & precision  & F  & precision  & F & precision  & F  \\
    \hline
    antV0 & 50  &50 & 60&50  &60&  70 \\
    antV1 & 60  &60 &50& 50  &60&  60 \\
    antV2 & 70  &90& 50&60  &60& 120 \\
    camelV0 & 70  &50& 70 &80   &110&  70 \\
    camelV1 & 60  &60&60&110   &70&  70\\
    ivy & 60 &60&  60&60   &60&  60 \\
    jeditV0 & 80  &80&80& 60   &90&  60 \\
    jeditV1 & 60  &70&80&  70   &80& 70 \\
    jeditV2 & 90  &80&60&70  &110&  80 \\
    log4j & 50 &70& 50&50   &80&  50 \\
    lucene & 80 &60& 70&60  &60&  70 \\
    poiV0 & 60  &60& 70 &60  &130&  80 \\
    poiV1 & 50 &50& 70 &50   &50&  110 \\
    synapse & 70  &60&50&  60   &50&  90 \\
    velocity & 60  &60&50& 60   &100&  60 \\
    xercesV0 & 60  &80&80& 60   &70& 80 \\
    xercesV1 & 80  &80&60& 60   &50&  80 \\
  \end{tabular}
  \caption{Number of evaluations for tuned learners, optimizing for precision and F-Measure.}
\label{tab:eval}
\end{table}

%%%%runtime for prec and F %%%%%%
\begin{table*}[!ht]
\renewcommand{\baselinestretch}{0.75}
\scriptsize
\centering
  \begin{tabular}{r|c c|c c|c c|c c|c c|c c }
    Datasets & \multicolumn{2}{c|}{Tuned\_Where} & \multicolumn{2}{c|}{Naive\_Where} & \multicolumn{2}{c|}{Tuned\_CART} & \multicolumn{2}{c|}{Naive\_CART} & \multicolumn{2}{c|}{Tuned\_RanFst} & \multicolumn{2}{c}{Naive\_RanFst}\\
    \hline
     & precision  & F    &  precision  & F  & precision  & F & precision  & F & precision  & F & precision  & F\\
    \hline
    antV0 & 95.47  &93.58  & 1.65 &1.39&  5.08 &3.52& 0.08 &0.08& 9.78 &9.89& 0.20 &0.17\\
    antV1 & 224.67 &186.95 & 3.03 &3.18&  6.52 &6.18& 0.12 &0.12&  14.13 &13.39& 0.25 &0.25\\
    antV2 & 644.99 &654.34 & 8.24 &8.08&  9.00 &8.79& 0.24 &0.18&  16.75 &27.56& 0.44&0.36\\
    camelV0 & 690.62 & 543.28 & 7.93 &9.65&  12.68 &17.00& 0.24 &0.28&  28.49 &22.52& 0.34&0.41\\
    camelV1 & 1596.77 & 1808.03 & 23.56 & 26.98 &  17.13 &31.92& 0.27 &0.28&  33.96 &37.00& 0.77&0.85\\
    ivy & 66.69 & 74.50 & 0.97 &1.18 &  4.26 &4.72& 0.07 &0.08& 8.89 &10.39& 0.19&0.21\\
    jeditV0 & 459.30&518.47 & 5.33 & 6.11 &  8.69 &7.9& 0.11 &0.10& 18.40 &14.32& 0.32&0.37\\
    jeditV1 & 421.56&576.29 & 6.59 & 6.89 &  9.05 &8.13& 0.12 &0.10& 17.93 &17.42& 0.36&0.34\\
    jeditV2 & 595.56&657.59 & 6.88 & 7.93 &  7.90 &10.34& 0.14 &0.15& 27.34 &20.20& 0.38&0.40\\
    log4j & 76.09 &123.48 & 1.33 & 1.59 &  2.60 &2.92& 0.06 &0.08& 9.69 &7.67& 0.15&0.17\\
    lucene & 236.45& 219.02& 2.60 & 3.68 &  6.07 &6.89 & 0.10 &0.12& 9.77 &13.06& 0.25&0.35\\
    poiV0 & 263.12 &314.53 & 3.92 & 4.82 &  7.42 &7.80& 0.09 &0.10& 25.86 &19.29& 0.28&0.32\\
    poiV1 & 398.33 & 446.05 & 6.94 & 7.55 &  9.31 &7.62& 0.13 &0.14& 12.67 &27.23& 0.29&0.36\\
    synapse & 144.09 & 138.75 & 1.85 & 1.83 &  3.88 &4.87& 0.07 &0.08& 8.13 &13.29& 0.19&0.17\\
    velocity & 184.10 &211.88 & 2.68 & 3.13 &  4.27 &5.51& 0.07 &0.10& 15.18 &11.58& 0.21&0.27\\
    xercesV0 & 136.87 & 178.49 & 1.98 & 2.02 &  9.17 &7.47& 0.10 &0.11& 14.17 &17.31& 0.22&0.28\\
    xercesV1 & 1173.92 & 1370.89 & 12.78 &14.42 &  10.47 &11.07 & 0.16 &0.19& 18.27 &25.27& 0.40&0.46\\
  \end{tabular}
  \caption{Runtime for tuned and default learners(in sec), optimizing for precision and F-Measure.}
  \label{tab:runtime}
\end{table*}

The number of evaluations and runtime used by our optimizers are shown in \tab{eval}
and \tab{runtime}.
WHERE's runtime are slower than CART and Random Forest since WHERE has yet to benefit from decades
of implementation experience with these older algorithms. For example, SciKitLearn's  CART and Random Forest
 make extensive use of an underlying C library whereas WHERE is a purely interpreted Python.

Looking over \tab{eval}, the general pattern is that 50 to 80 evaluations suffice for finding the tuning
improvements reported in this paper. 
50 to 80 evaluations are  much fewer than our pre-experimental intuition.
Prior to this paper, the authors have conducted numerous explorations of evolutionary algorithms
for search-based SE applications~\cite{krall15,krall15:hm,fea02a,me07f,Green}. Based
on that work, our expectations were that non-parametric evolutionary optimization would
take thousands, if not millions, of evaluations of candidate tunings. This turned out not
to be that case. By comparing the runtime of tuned and default learners shown in \tab{runtime}, we notice
that the actual tuning time for most data sets is not extremely long.

Hence, we answer RQ5 as ``no'': tuning is so fast that
it could (and should) be used by anyone using defect predictors. 
%The possible reason that tuning is so fast is that the searching space for defect prediction is not complicated, which might result from the tuning range set for each parameter in \tab{parameters}.  

As to why DE can tune defect predictors so quickly, that is an
open question. One possibility is that the search space within
the control space of these data miners has  many accumulative effects such that one
decision can cascade into another (and the combination of decisions
is better than each separate  one). DE would be a natural  tool for reasoning
about such ``cascades'', due to the way it mashes candidates together,
then inserts the result back into the frontier (making them available
for even more mashing at the next step of the inference).

\subsection{RQ6: Should we use ``off-the-shelf'' Tunings?}\label{sect:variance}
 
 In \fig{features}, we show how tuning selects the optimal values for tuned parameters. For space limitation, only four parameters from WHERE learner are selected as representatives and all the others can be found in our online support documents (https://goo.gl/aHQKtU).
 Note that
 the tunings learned were different in different data sets and for different goals.
Also, the tunings learned by DE
were often very different to the default (the default values for {\em threshold}, {\em infoPrune}, {\em min\_Size} and {\em wriggle} are $0.5$, $0.33$, $0.5$ and $0.2$, respectively). That is, to achieve the performance improvements seen in the paper,
the default tuning parameters required a wide range of adjustments.

Hence, we answer RQ6 as ``no'' since, to achieve the improvements seen in this paper, tuning has to be repeated whenever the goals or data
sets are changed. Given this requirement to repeatedly run tuning, it is fortunate that (as shown above)
tuning is so easy and so fast (at least for defect predictors from static code attributes).

\section{Reliability and Validity}\label{sect:construct}

{\em Reliability} refers to the consistency of the results obtained
from the research.  For example,   how well independent researchers
could reproduce the study? To increase external
reliability, this paper has taken care to either  clearly define our
algorithms or use implementations from the public domain
(SciKitLearn). Also, all the data used in this work is available
on-line in the PROMISE code repository and all our algorithms
are on-line at github.com/ai-se/where.

{\em External validity} checks if the results are of relevance
for other cases, or can be generalized from samples
to populations.  
The examples of this paper  only relate to precision, recall, and the F-measure
but the general principle (that the search bias changes the search conclusions)  holds for any set of goals. 
Also,
the tuning results shown here only came from one  software analytics task 
(defect prediction from static code attributes).
There are many other kinds of software analytics tasks 
(software development effort estimation, social network mining,
detecting duplicate issue reports, etc) and the implication of this
study for those tasks is unclear. 
However,  those other tasks often use the same kinds of learners
explored in this paper so it is quite possible that
the conclusions of this paper apply to other SE analytics tasks as well. 

%That said, there exist some class of data mining papers for which
%tuning may not be required. Consider  Le Goues et al.'s 2012
%ICSE paper that used a evolutionary program to learn
%repairs to code~\cite{leGoues12}. The performance criteria
%in that paper was ``can we fix any of the known bugs?''. Note
%that this criteria is a ``{\em competency}'' statement, and
%not a ``{\em better than}'' statement (the difference being that
%one is 
%``can do'' and the other is ``can do better''). For such
%competency claims, tuning is not necessary. However, as soon
%as {\em better than} enters the performance criteria then this
%becomes a race between competing methods. In such a race,
%it is unfair to hobble one competitor with poor tunings.

\section{Conclusions}

Our exploration of the six research
questions listed in the introduction
show that when learning defect predictors for static code
attributes,   analytics without parameter tuning are considered {\em harmful} and {\em misleading}:
\bi
\item Tuning improves the performance scores of a predictor.
That improvement is usually positive (see \fig{deltas}) and sometimes
it can be quite   dramatic (e.g. precision changing from 0 to 60\%). \item 
Tuning changes conclusions on what learners are better than others.
Hence, it is time to revisit numerous prior publications of our own~\cite{me07b}
and others~\cite{lessmann2008benchmarking,hall11}.
\item
Also,
tuning changes conclusions on what factors are most important in software development.
Once again, this means that old papers may need to be revised including those
some of our own~\cite{me02k} and others~\cite{bell2013limited,rahman2013how,Moser:2008,zimmermann2007predicting,herzig2013predicting}. 
\ei
As to future work, it is now important
to explore the implications of these
conclusions to other kinds of software analytics.
 This paper has investigated  {\em some} learners using {\em one}  optimizer. Hence, we can make
no claim that DE is the {\em best} optimizer for {\em all} learners.
Rather, our point is that there exists at least some learners
whose performance can be dramatically improved by 
at least one simple optimization scheme.  We hope that this work inspires
much future work as this community develops and debugs best practices for tuning
software analytics.

Finally, on a more general note, we point out
that 
F\"{u}rnkranz~\cite{furnkranz05} says data mining is inherently a multi-objective optimization
problem that seeks the smallest model with the highest performance, 
that generalizes best for
future examples (perhaps learned in minimal time using the least amount of data).
In this view, we are using DE to optimize an optimizer. Perhaps a better approach might be
to dispense with the separation of ``optimizer'' and ``learner'' and combine them both
into one system that learns how to tune itself as it executes. If this view is useful,
then instead of adding elaborations to data miners (as done in this paper, or by researchers
exploring hyper-heuristics~\cite{jia2013learning}), it should be possible to radically simplify optimization and data
mining with a single system that rapidly performs both tasks.

\section*{Acknowledgments}
The work has partially funded by a National Science Foundation CISE CCF award \#1506586.
 
\vspace*{0.5mm}

\bibliographystyle{elsarticle-num}
% \balance
\bibliography{main}

\end{document}